# Phonon scattering limited mobility in the representative cubic perovskite semiconductors $SrGeO_3$, $BaSnO_3$ and $SrTiO_3$

Christian A. Niedermeier,[1,*] Yu Kumagai,[1] Keisuke Ide,[1] Takayoshi Katase,[1] Fumiyasu Oba,[1,2,] Hideo Hosono,[1,2] and Toshio Kamiya[1,2]

**Corresponding author: c-niedermeier@mces.titech.ac.jp*

[1]Laboratory for Materials and Structures, Tokyo Institute of Technology, Yokohama 226-8503, Japan

[2]Materials Research Center for Element Strategy, Tokyo Institute of Technology, Yokohama 226-8503, Japan

**Abstract:** Cubic perovskite oxides are emerging high mobility transparent conducting oxides (TCOs), but Ge-based TCOs had not been known until the discovery of metastable cubic $SrGeO_3$. $0.5 \times 0.4 \times 0.2\text{-mm}^3$ large single crystals of the cubic $SrGeO_3$ perovskite were successfully synthesized employing the high-pressure flux method. The phonon spectrum is determined from the IR optical reflectance and Raman scattering analysis to evaluate the electron transport governed by optical phonon scattering. A calculated room-temperature mobility on the order of $3.9\times10^2$ $cm^2V^{-1}s^{-1}$ is obtained, identifying cubic $SrGeO_3$ as one of the most promising TCOs. Employing classical phonon theory and a combined experimental-theoretical approach, a comprehensive analysis of the intrinsic electron mobility in the cubic perovskite semiconductors $SrGeO_3$, $BaSnO_3$ and $SrTiO_3$ is provided based on the magnitude of polarization and eigenfrequency of optically active phonons.

# I. INTRODUCTION

TCOs are emerging materials combining high electrical conductivity and optical transparency in the visible spectral region, and present important applications in optoelectronic devices. Since the discovery of the high-mobility two-dimensional electron gas in oxide heterostructures,[1] perovskites became particularly attractive for advanced semiconductor technology as they facilitate a new design of future electronic devices due to the unique physics occurring at the interfaces of hetero-junctions.[2] The recent reports of the remarkably high electron mobility at cryogenic temperatures in La:$SrTiO_3$ epitaxial thin films (>50000 $cm^2V^{-1}s^{-1}$)[3,4] and at room temperature (RT) in La:$BaSnO_3$ single crystals (320 $cm^2V^{-1}s^{-1}$)[5,6] sparked significant interest in understanding the fundamentals of the intrinsic electron transport in perovskite semiconductors and their application in heterostructures.[7–9]

The discovery of the first Ge-based TCO, cubic perovskite $SrGeO_3$,[10] invites the exploration of a new, promising emerging material for application in electronics. The large conduction band (CB) dispersions in the main group element cubic perovskites $SrGeO_3$ and $BaSnO_3$ originate from the regular octahedral coordination of the $Ge^{4+}$ ($Sn^{4+}$) cation to $O^{2-}$ anions with the point symmetry $O_h$.[11,12] This coordination symmetry prohibits the hybridization of Ge 4s (Sn 5s) and O 2p orbitals at the $\Gamma$ point in the Brillouin zone. Thus, the resulting nonbonding state at the $\Gamma$ point and the antibonding states at the boundaries of the Brillouin zone have energy levels separated by several eV. The large CB curvatures yield small effective masses of $SrGeO_3$ (0.22 $m_e$) and $BaSnO_3$ (0.19 $m_e$),[13,14] where $m_e$ denotes the electron rest mass.

The relatively large static dielectric permittivity of perovskites is a benefical material property which screens the electron-charge interaction and promotes the electron transport. It results from the strongly ionic character of the chemical bonds and the ion displacement from the equilibrium lattice sites. These thermally activated lattice vibrations, known as phonons, may account for a large ionic polarization even when the material is composed of only main group elements. The static dielectric constants of $SrGeO_3$ ($18 \pm 1$) and $BaSnO_3$ (20,[15] $20 \pm 2$,[16,17] $14.8 \pm 0.5$[18]) are about 2-3 times larger than those of the well-established TCOs like ZnO ($\varepsilon_s^{\|c}$ =7.4, $\varepsilon_s^{\perp c}$ = 8.5),[19] $In_2O_3$ ($\varepsilon_s$ = 8.5)[20] and $SnO_2$ ($\varepsilon_s^{\|c}$ = 14, $\varepsilon_s^{\perp c}$ = 9.0).[21] Although recent Berry phase calculations have revealed that electron transfer has a large contribution to such large dielectric polarizations, their origins have not yet been discussed in relation to phononic polarizations experimentally. Due to the large polarization, the RT electron mobility in substitutionally doped perovskite oxides is governed by electron-phonon interactions rather than ionized impurity scattering,[22,23] and the dielectric screening of ionized impurities supports a high electron mobility even at enhanced carrier concentrations.[14,24] In contrast to the majority of TCOs with a ($n$-1)$d^{10}ns^0$ electronic configuration ($n$ is the principal quantum number), transition metal-based semiconductors such as $SrTiO_3$ and $TiO_2$ have rather localized unoccupied wave functions originating from the Ti 3d orbitals and a large effective mass. However, these semiconductors may nevertheless exhibit fairly good electrical transport properties. In this case, the large dielectric screening ($SrTiO_3$, $\varepsilon_s$ = 310;[25] and $TiO_2$, $\varepsilon_s^{\parallel c}$ = 173, $\varepsilon_s^{\perp c}$ = 89)[26] is seen as a key property for promoting the electrical conductivity.[27]

This work employs a combined experimental-theoretical approach to provide a fundamental understanding of how the ionic polarization, decoded into the contribution of separate phonon modes, impacts the intrinsic electron transport in cubic perovskite semiconductors. $SrTiO_3$ and $BaSnO_3$ have been well characterized, and recently, also large-size single $BaSnO_3$ crystals could be grown from the melt[28] and hydrothermal methods.[29] In contrast, up to present, the cubic perovskite $SrGeO_3$ (Fig. 1a) has remained widely unexplored due to the difficulty in the synthesis, requiring high-pressure conditions.[30] So far, only a $0.1 \times 0.08 \times 0.08$-$mm^3$ small single crystal was obtained employing the high-pressure sintering of a polycrystalline sample.[31] We report the successful single crystal growth of cubic $SrGeO_3$ employing the high-pressure flux method. The phonon spectrum of cubic $SrGeO_3$ was determined, which allows for a quantitative analysis of the intrinsic mobility, in comparison with $SrTiO_3$ and $BaSnO_3$. We conclude that the particular combination of properties, a large polarization of the hardest longitudinal optical (LO) phonon mode *and* a small effective mass, would be effective for the design of electronic materials with an exceptionally high-RT electron mobility.

## II. MATERIALS AND METHODS

### A. Experimental

Cubic $SrGeO_3$ single crystals were grown using a belt-type high-pressure apparatus.[32] The precursor powder, monoclinic $SrGeO_3$ (spacegroup *C2/c*), was prepared by calcination of stoichiometric quantities of $SrCO_3$ and $GeO_2$ at 1100 $^{o}$C for 12 h. The monoclinic $SrGeO_3$ powder (100 mg) was mixed in the 1:7 molar ratio with a $SrCl_2$/NaCl flux of the molar composition 3:2. The eutectic point in the $SrCl_2$-NaCl system lies at 544 $^{o}$C and 53 mol% $SrCl_2$.[33] $SrCl_2$ and NaCl were dehydrated at 350 $^{o}$C prior to the mixing. The powder mixture was filled into a Au capsule, pressed and inserted into a NaCl/$ZrO_2$ cell of the weight ratio 9:1 with a carbon heater. A 5-GPa pressure was applied and the specimen was annealed at 1100 $^{o}$C for 4 hours, then slowly cooled down to 800 $^{o}$C within 18 h, at which point the heating was turned off and the specimen was water-cooled to room temperature. After washing out the flux with deionized water, yellowish transparent cubic $SrGeO_3$ ($Pm\overline{3}m$) single crystals of up to 0.5 $\times$ 0.4 $\times$ 0.2 mm$^3$ size and naturally grown 100 facets were obtained (Fig. 1b). Colorless, orthorhombic $\beta$-$SrGe_2O_5$ (*Pnma*) single crystals were obtained as side product.[34] We also prepared dense, cubic $SrGeO_3$ and $La_{0.02}Sr_{0.98}GeO_3$ polycrystalline pellets of 5 mm diameter by sintering the precursor powder in a BN capsule at 1100 $^{o}$C for 2 h at 5-GPa pressure. The pellets were cleaved and the surface was mirror-polished for optical reflectance measurements.

The 100 out-of-plane high-resolution X-ray diffraction (HR-XRD) pattern of cubic $SrGeO_3$ single crystals and the rocking curve of the 200 peak were recorded using a Rigaku SmartLab diffractometer equipped with a monochromatic Cu $K_{\alpha 1}$ X-ray source (1.5406 Å) and utilizing parallel beam optics. Reflection ellipsometry spectra (Jobin Yvon UVISEL) were recorded from 0.6–5.0 eV. Specular reflection IR spectra (Hitachi U-4100) were recorded from 1.5–5.0 eV and using vacuum Fourier-transform IR spectroscopy (Bruker Vertex 70v) from 100–8000 cm$^{-1}$ wavenumber. Al and Au mirrors were used as 100% reflectance standards, respectively. For measuring the nonpolarized Raman spectrum of cubic $SrGeO_3$ single crystals, we used a Horiba LabRAM HR system in the backscattering configuration employing the 457 nm wavelength of a Nd:$YVO_4$ laser (Cobolt Twist 50) and a liquid $N_2$-cooled CCD detector.

### B. Computational

We performed the theoretical calculations using hybrid density functional theory (DFT) and the projector augmented wave method[35] as implemented in the Vienna Ab initio Simulation Package (VASP).[36,37] As the cubic perovskite has no degree of freedom for internal atomic coordinates, only the lattice parameter *a* was relaxed by using the Perdew-Burke-Ernzerhof (PBE0) hybrid functional.[38] A plane wave cutoff of 500 eV and a $\Gamma$-centered 6 x 6 x 6 *k*-mesh were used. The relaxed ground state cubic $SrGeO_3$ structure has the calculated lattice parameter of 3.80 Å, which is in excellent agreement to the experimental value at room temperature, 3.7978 Å.[31] The electronic band structure was calculated employing maximally localized Wannier functions using the code Wannier90.[39] We employed the PBE0 functional because it gave a reasonable band gap for related oxides, $GeO_2$ (tetragonal, *P4₂/mnm*, $E_{\mathrm{g}}^{\mathrm{calc}}$ = 4.55 eV, $E_{\mathrm{g}}^{\mathrm{exp}}$= 4.68 eV)[40] and $BaSnO_3$ ($E_{\mathrm{g}}^{\mathrm{calc}}$ = 3.17 eV, $E_{\mathrm{g}}^{\mathrm{exp}}$= 2.95 eV).[24]

We used the phonopy code[41] and adopted the experimental lattice parameters at room temperature since they largely affect the phonon frequencies. For the exchange-correlation functional, we adopted the generalized gradient approximation with a very high accuracy of total energy criteria for self-consistent convergence, $10^{-8}$ eV. 2 x 2 x 2 40-atom supercells were used along with a plane wave cutoff of 400 eV and a $\Gamma$-centered 4 x 4 x 4 *k*-mesh.

## III. CRYSTAL STRUCTURE, ELECTRONIC STRUCTURE AND OPTICAL CHARACTERIZATION OF CUBIC $SrGeO_3$

The 100 out-of-plane HR-XRD pattern of the $SrGeO_3$ single crystal confirms the cubic structure and a 3.7980-Å lattice parameter is obtained using Bragg's law (Fig. 1c). The narrow, 0.023° wide full-width at half maximum (FWHM) of the rocking curve of the 200 peak indicates a low degree of mosaicity of the single crystal (Fig. S1). The PBE0-calculated band structure of the cubic perovskite $SrGeO_3$ shows a 2.7-eV indirect band gap of the CB minimum at the *Γ* point and the valence band (VB) maximum at the *R* point of the first Brillouin zone (Fig. 1d). The effective mass of 0.22 $m_e$ is determined from the CB curvature at the $\Gamma$ point[42]

$$m_e^* = \hbar^2 \left(\frac{d^2E}{dk^2}\right)^{-1} \quad (1)$$

where $\hbar$ is the reduced Planck constant and $k$ is the reciprocal space vector. Using a cleaved and mirror-polished cubic $SrGeO_3$ polycrystalline pellet, we obtained the reflection ellipsometry spectrum (Fig. 1e) to evaluate the absorption coefficient $\alpha$. Note that the absorption below the optical band gap is a measurement artefact due to the residual surface roughness which causes diffuse scattering or trace sub-gap absorption from in-band gap defects. The linear fits to the $\alpha^{1/2}$ and $\alpha^2$–photon energy plots in the vicinity of the absorption edge allow for the determination of 2.7-eV indirect and 3.4-eV direct optical band gaps, respectively, in agreement with the specular reflectance edge and the calculated values (Fig. 1f).

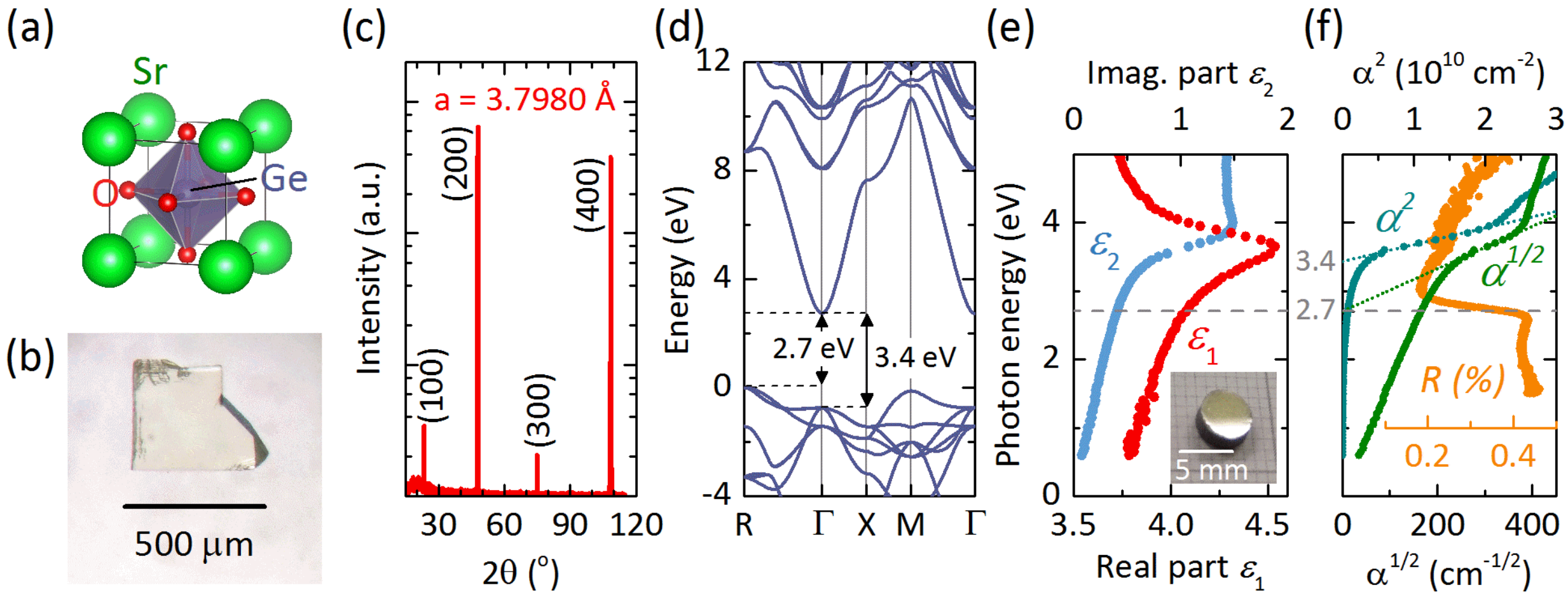

Fig. 1: (a) The cubic perovskite $SrGeO_3$ crystal structure. (b) Optical micrograph of a cubic $SrGeO_3$ single crystal. (c) The 100 out-of-plane HR-XRD pattern of a cubic $SrGeO_3$ single crystal with the intensity plotted on a logarithmic scale. (d) PBE0-calculated band structure along high symmetry directions in the first Brillouin zone indicates 2.7 eV indirect and 3.4 eV direct band gaps and a large CB dispersion. (e) Reflection ellipsometry spectrum for the evaluation of (f) the absorption coeffient data $\alpha^{1/2}$ and $\alpha^2$ indicate optical band gaps in agreement with the specular reflectance *R* edge and the calculated values.

## IV. PHONON SPECTRA OF CUBIC PEROVSKITES

### A. IR dispersion function of cubic $SrGeO_3$

To investigate the intrinsic electron mobility in cubic $SrGeO_3$, we first determine the phonon spectrum. Materials crystallizing in the cubic perovskite structure (one $SrGeO_3$ chemical unit) have a total of 15 phonon modes;[43] 3 acoustic modes, 4 pairs of doubly degenerate transverse optical (TO) modes and 4 longitudinal optical (LO) phonon modes. Only three of the four triplets of optical modes are IR active, having a net ionic dipole moment which contributes to the macroscopic static polarization. Thus, we construct the optical dispersion model using a Lorentz oscillator[44] for each of the three IR active TO phonon modes,

$$\varepsilon(\omega) = \varepsilon_\infty + \sum_\mu \frac{\varepsilon_{t_\mu} \omega_{t_\mu}^2}{\omega_{t_\mu}^2 - \omega^2 - i\gamma_\mu \omega} \tag{2}$$

where $\varepsilon_\infty$ is the high-frequency dielectric constant, $\omega_{t_\mu}$ is the TO phonon eigenfrequency and $\gamma_\mu$ is the damping frequency, both given as angular frequencies. The contribution of the individual phonon modes to the static polarization is given by the TO phonon oscillator strength $\varepsilon_{t_\mu}$. Following the classical theory of polarization waves in crystals,[45] the TO phonon oscillator strength is given by

$$\varepsilon_{t_\mu} = \frac{(\sum_i e_i^* x_{t_\mu i})^2}{\omega_{t_\mu}^2 \varepsilon_0 V \sum_i M_i x_{t_\mu i}^2} \tag{3}$$

where $\varepsilon_0$ is the vacuum permitivity and $V$ is the unit-cell volume. The ionic polarization results from the displacement $x_{t_\mu i}$ of ions with the Born effective charge $e_i^*$ and the atomic mass $M_i$. This classical theory holds for the harmonic vibration regime, in which ion displacements are sufficiently small, such that the induced polarization can be described as a linear function of $x_{t_\mu i}$. In fact, the displacement of a certain ion would also induce a polarization on other ions as well, giving a rather complicated description of the total polarization.

For the determination of the dispersion parameters in Eq. (2), we fit the specular IR reflectance spectrum of a mirror-polished cubic $SrGeO_3$ polycrystalline pellet (Fig. 2a). In addition, we measured the Raman spectrum for a cubic $SrGeO_3$ single crystal. The first-order Raman peaks provide the LO phonon frequencies, which are given by the maxima of the loss function $-\mathrm{Im}(\varepsilon^{-1})$.[46] According to theory, cubic perovskites shall not show any first-order Raman active modes,[43] but we observe the $LO_2$ and $LO_3$ phonons at 493-cm$^{-1}$ and 742-cm$^{-1}$ wavenumber, respectively, as well as their overtones at 995 cm$^{-1}$ and 1484 cm$^{-1}$ (Fig. 2b). The shoulder peak at 695 cm$^{-1}$ may relate to the $LO_1$ + $LO_2$ overtone. The $LO_1$ mode can not be identified due to its negligibly small polarization. The Raman spectrum of cubic $BaSnO_3$ is strikingly similar to that of cubic $SrGeO_3$, and a large scattering intensity of the $LO_2$ and $LO_3$ phonons is observed.[16] The large Raman response of LO modes has been attributed to resonant multi-phonon scattering known to occur in materials of large lattice polarizability.

The IR reflectance spectrum was fitted by first constraining the $LO_2$ and $LO_3$ frequencies to 493 cm$^{-1}$ and 792 cm$^{-1}$, respectively, and determining the high-frequency dielectric constant $\varepsilon_\infty$= 4.2 $\pm$ 0.2 from the position of the reflectance edge near a frequency of 760 cm$^{-1}$. Then, all dispersion parameters were allowed to vary for the final fitting to obtain a reliable dispersion model which can very well reproduce both, the IR reflectance spectrum and the frequencies of the first-order $LO_2$ and $LO_3$ Raman peaks. We do not attempt to fit the Raman scattering intensity, which is a complicated function of wavenumber that cannot be simply described by considering only the loss function $-\mathrm{Im}(\varepsilon^{-1})$.[47] The fitting was performed using the RefFIT program[48] and the obtained dispersion parameters are given in Table 1. The cubic $SrGeO_3$ static dielectric constant $\varepsilon_s$ = 18 $\pm$ 1 is obtained from the sum of the high-frequency dielectric constant and all three TO phonon oscillator strengths $\varepsilon_{\mathrm{t}_\mu}$. The largest contribution to the total static polarization originates from the $TO_2$ mode ($\varepsilon_{\mathrm{t}_2} = 11.2$).

The result of our dispersion model contradicts the finding of the apparently very high static dielectric constant ($\varepsilon_s$ = 114) and largely polar $TO_1$ mode ($\varepsilon_{\mathrm{t}_1}$ = 100.5) in cubic $SrGeO_3$ reported by Grzechnik.[49–51] Our experimental results are further supported by the agreement with the theoretical calculation of the cubic $SrGeO_3$ static dielectric constant $\varepsilon_s^{\mathrm{calc}} =$ 21 using density functional pertubation theory.[52] Moreover, it is reasonable that the cubic $SrGeO_3$ dispersion parameters are comparable to those reported for $BaSnO_3$ ($\varepsilon_s$ = 20,[15] 20 $\pm$ 2,[16,17] 14.8 $\pm$ 0.5[18]), due to the similarities in the chemical bonding and electronic structure. We also calculated the high-frequency dielectric constant $\varepsilon_\infty^{\mathrm{calc}} =$ 4.9 of cubic $SrGeO_3$, which is slightly larger as compared to the experimental value.

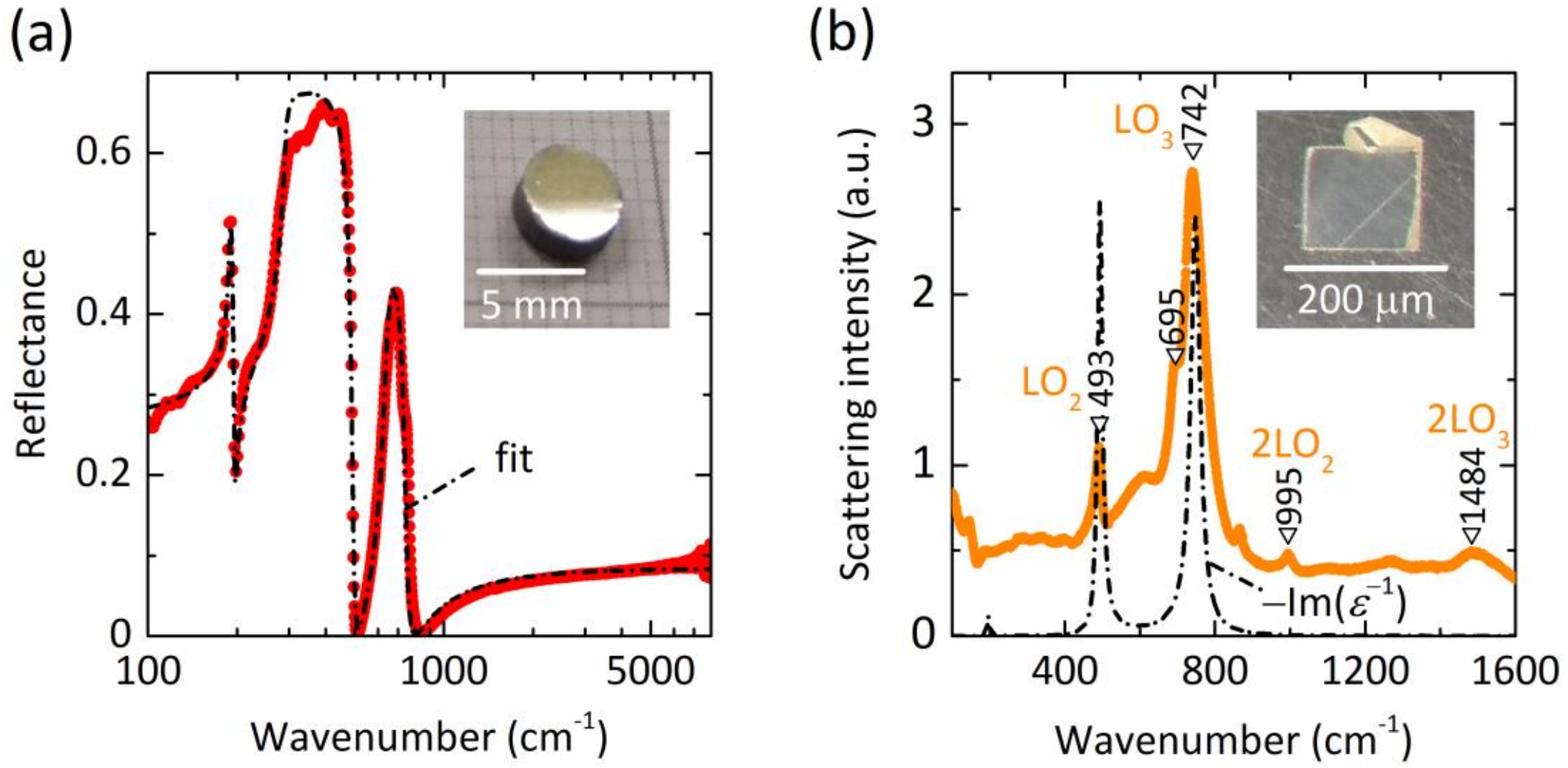

Fig. 2: (a) IR reflectance spectrum of the polished cubic $SrGeO_3$ polycrystalline pellet. (b) Raman-scattering spectrum of a cubic perovskite $SrGeO_3$ single crystal. The first-order Raman peaks at 493 $cm^{-1}$ and 742 $cm^{-1}$ provide the $LO_2$ and $LO_3$ phonon frequencies in the dispersion model, as given by the maxima of the loss function $-\mathrm{Im}(\varepsilon^{-1})$.

Table 1: TO phonon frequency $\omega_{\mathrm{t}_\mu}$, damping frequency $\gamma_\mu$ and oscillator strength $\varepsilon_{\mathrm{t}_\mu}$ describing the $SrGeO_3$, $BaSnO_3$ and $SrTiO_3$ IR dielectric functions using three Lorentz oscillators (Eq. 2). The high-frequency dielectric constants are $\varepsilon_\infty^{\mathrm{SrGeO3}} = 4.2 \pm 0.2$, $\varepsilon_\infty^{\mathrm{BaSnO3}} = 4.3$[16] and $\varepsilon_\infty^{\mathrm{SrTiO3}} = 5.2$.[25] Values in brackets denote calculation results using the PBE functional. The theoretical TO phonon oscillator strength $\varepsilon_{\mathrm{t}_\mu}$ is obtained from the phonon calculation results of the ion displacement vectors and Born effective charges (Eq. (3)).

| Material | Eigenfrequency $\omega_{\mathrm{t}_\mu}$ (cm$^{-1}$) | Energy $\hbar\omega_{\mathrm{t}_\mu}$ (meV) | Damping frequency $\gamma_\mu$ (cm$^{-1}$) | Oscillator strength $\varepsilon_{\mathrm{t}_\mu}$ - |
|---|---|---|---|---|
| $SrGeO_3$ | 189 (180) | 23 | 4.0 | 1.8 (3.1) |
| | 291 (270) | 36 | 7.6 | 11.2 (9.8) |
| | 634 (671) | 79 | 39 | 0.83 (1.4) |
| $BaSnO_3$[a)] | 135 (136) | 17 | 10 | 3.8 $\pm$ 2 (3.8) |
| | 246 (231) | 30 | 18 | 10.8 (13.8) |
| | 629 (641) | 78 | 26 | 0.76 (1.0) |
| $SrTiO_3$[b)] | 88 (81) | 11 | 26 | 299 (288) |
| | 178 (173) | 22 | 6.1 | 3.6 (21) |
| | 546 (561) | 68 | 27 | 1.9 (2.9) |

[a)]References[16,17]; [b)]Reference[25]

### B. LO phonon oscillator strength

The LO phonon oscillator strength $\varepsilon_{\mathrm{l}_\nu}$ is calculated from the dispersion model parameters, after the expansion of the longitudinal mode displacements $x_{\mathrm{l}_\nu}$ in terms of those of the transverse modes $x_{\mathrm{t}_\mu}$,[53] according to

$$\varepsilon_{\mathrm{l}_\nu} = \varepsilon_\infty^2 \left( \omega_{\mathrm{l}_\nu}^2 \sum_\mu \frac{\varepsilon_{\mathrm{t}_\mu} \omega_{\mathrm{t}_\mu}^2}{(\omega_{\mathrm{l}_\nu}^2 - \omega_{\mathrm{t}_\mu}^2)^2} \right)^{-1} \tag{4}$$

with the results given in Table 2. The matrix elements of Eq. (4) describe the coupling strength between TO and LO modes (Table 3).

Table 2: LO phonon frequency $\omega_{\mathrm{l}_\nu}$, LO phonon oscillator strength $\varepsilon_{\mathrm{l}_\nu}$ for $SrGeO_3$, $BaSnO_3$ and $SrTiO_3$. Values in brackets denote calculation results using the PBE functional. The theoretical LO phonon oscillator strength $\varepsilon_{\mathrm{l}_\nu}$ is obtained from the phonon calculation results of the ion displacement vectors and Born effective charges (Eq. (3)). The polaron coupling constant $\alpha_\nu$ and RT relaxation time $\tau_\nu^{\mathrm{RT}}$ is calculated according to Eqs. (5) and (6), respectively.

| Material | Eigenfrequency | Energy | Oscillator strength | Coupling constant | RT relax. time |
|---|---|---|---|---|---|
| | $\omega_{\mathrm{l}_\nu}$ | $\hbar\omega_{\mathrm{l}_\nu}$ | $\varepsilon_{\mathrm{l}_\nu}$ | $\alpha_\nu\left(\frac{m_\mathrm{e}}{m_\mathrm{e}^*}\right)^{1/2}$ | $\tau_\nu^{\mathrm{RT}}$ |
| | ($cm^{-1}$) | (meV) | - | - | ($10^{-14}$ s) |
| $SrGeO_3$ | 196 (191) | 24 | $0.047 \pm 0.01$ (0.09) | 0.062 | 72 |
| | 493 (426) | 61 | $1.4 \pm 0.2$ (1.9) | 1.2 | 7.5 |
| | 748 (791) | 93 | $1.8 \pm 0.2$ (1.4) | 1.2 | 17 |
| $BaSnO_3$ | 146 [a)] (148) | 18 | $0.11 \pm 0.07$ (0.15) | 0.17 | 25 |
| | 425 [a)] (370) | 53 | $1.9 \pm 0.1$ (2.1) | 1.6 | 4.4 |
| | 720[a)] (726) | 89 | $1.4 \pm 0.1$ (1.3) | 0.92 | 24 |
| $SrTiO_3$ | 174 (154) | 22[b)] | 0.02 (0.08) | 0.017[b)] | 89 |
| | 461 (438) | 57[b)] | 0.94 (1.7) | 0.54[b)] | 4.8 |
| | 816 (807) | 101[b)] | 4.2 (1.6) | 1.8[b)] | 2.8 |

[a)]Reference[16]; [b)]cf. reference[54].

Table 3: The matrix elements of Eq. (4) describe the strength of coupling between TO and LO phonon modes, from which the LO phonon oscillator strength $\varepsilon_{l_\nu}$ is obtained.

| Material | | $\omega_{l_\nu}^2 \frac{\varepsilon_{t_\mu}\omega_{t_\mu}^2}{\left(\omega_{l_\nu}^2-\omega_{t_\mu}^2\right)^2}$ | | | $\omega_{l_\nu}^2 \sum_\mu \frac{\varepsilon_{t_\mu}\omega_{t_\mu}^2}{\left(\omega_{l_\nu}^2-\omega_{t_\mu}^2\right)^2}$ |
|---|---|---|---|---|---|
| | $\nu$ | $\mu = 1$ | $\mu = 2$ | $\mu = 3$ | |
| $SrGeO_3$ | 1 | 364 | 17.1 | 0.1 | 381 |
| | 2 | 0.4 | 9.2 | 3.2 | 12.8 |
| | 3 | 0.1 | 2.4 | 7.5 | 10.0 |
| $BaSnO_3$ | 1 | 154 | 9.1 | 0.1 | 163 |
| | 2 | 0.5 | 8.1 | 1.3 | 9.9 |
| | 3 | 0.1 | 1.6 | 11.7 | 13.4 |
| $SrTiO_3$ | 1 | 137 | 1372 | 0.2 | 1510 |
| | 2 | 11.6 | 0.7 | 16.4 | 28.7 |
| | 3 | 3.5 | 0.2 | 2.8 | 6.5 |

### C. Comparison of the cubic perovskites $SrGeO_3$, $BaSnO_3$ and $SrTiO_3$

In Fig. 3 we provide the IR dispersion functions of cubic $SrGeO_3$ in comparison with $BaSnO_3$ and $SrTiO_3$, which were determined by optical analyses of single crystals.[16,17,25] The maxima of the imaginary part $\varepsilon_2$ of the dielectric function provide the TO phonon frequencies $\omega_{t_\mu}$. The maxima of the loss function –Im($\varepsilon^{-1}$) provide the LO phonon frequencies $\omega_{l_\nu}$, which correspond to the zeros of the real part $\varepsilon_1$. The peak heigths of the TO and LO phonons indicate the magnitude of polarization, but it must be considered that the peak broading due to damping ($\gamma_\mu > 0$) is taken into account. The phonon spectra of $SrGeO_3$ and $BaSnO_3$ are very similar, and both materials show large polarizations for the $TO_2$, $LO_2$ and $LO_3$ modes (Fig. 3b). However, the $SrGeO_3$ eigenfrequencies of these modes are noticeably shifted to higher frequencies. The $SrTiO_3$ phonon spectrum is distinctly different and shows a remarkably large polarization of the $TO_1$ and $LO_3$ modes.

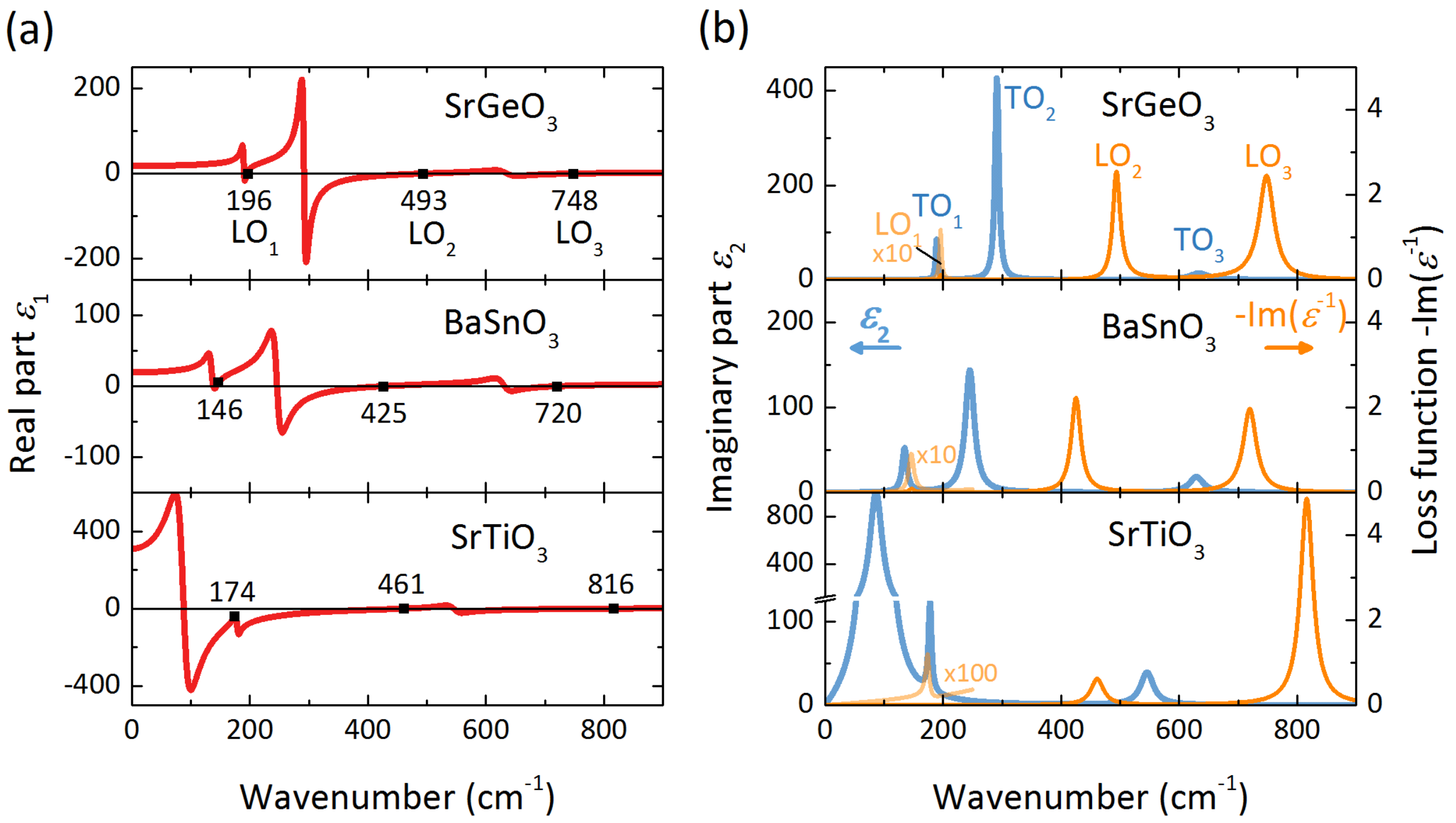

Fig. 3: (a) Real part of $\varepsilon_1$ of the dielectric function of cubic $SrGeO_3$, $BaSnO_3$ and $SrTiO_3$. The LO phonon frequencies correspond to the zeros of $\varepsilon_1$. (b) The maxima of the imaginary part $\varepsilon_2$ of the dielectric function and the loss function $-\text{Im}(\varepsilon^{-1})$ correspond to the frequency of TO and LO phonon modes, respectively. The peak heigths of the TO and LO phonons indicate the magnitude of polarization, but it must be considered that the peak broading due to damping ($\gamma_\mu > 0$) is taken into account.

To understand the differences in the phonon spectra between cubic $SrGeO_3$ and $BaSnO_3$, and $SrTiO_3$ on the other hand, we calculated the phonon band structures and displacement vectors of all eigenmodes (Fig. 4). The $SrGeO_3$ and $SrTiO_3$ phonon band structures show imaginary frequencies at the Brillouin zone boundaries, because at ambient pressure and absolute-zero temperature, the thermodynamically stable crystal structures are not cubic ($Pm\overline{3}m$), but monoclinic ($C2/c$) and tetragonal ($I4/mcm$),[55] respectively. The 3 IR active TO eigenmodes in $SrGeO_3$ and $BaSnO_3$ are very similar, and one schematic is given to illustrate these modes in both materials. There are different displacement vectors for the following ions: $A^{2+}$ and $B^{4+}$ cations, and $O_{||}^{2-}$ and $2{\times}O_{\perp}^{2-}$ anions with displacements vectors parallel and perpendicular to the axis of the B—O bond, respectively.

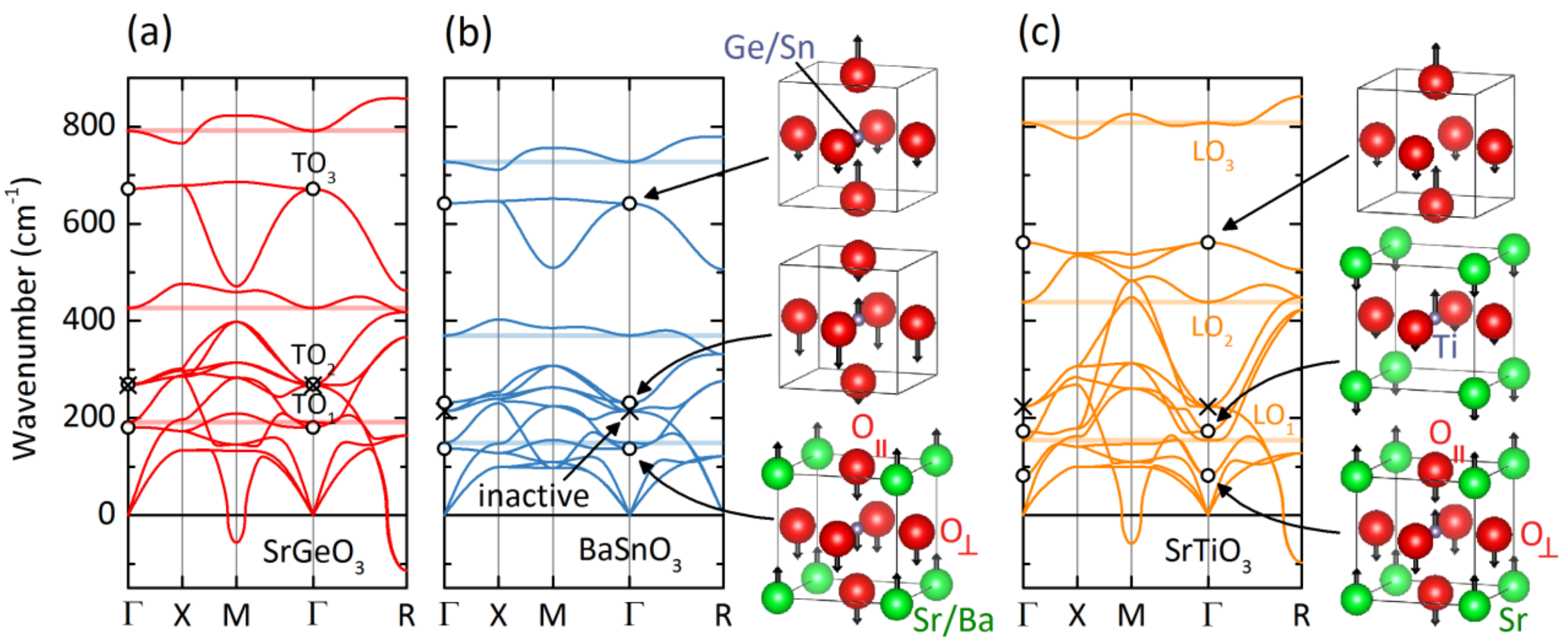

Fig. 4: Calculated phonon band structures of cubic $SrGeO_3$, $BaSnO_3$ and $SrTiO_3$ obtained by density functional perturbation theory. The 3 IR active TO modes at the $\Gamma$ point of the first Brillouin zone (empty circles) are schematically illustrated based on the calculated ion displacement vectors. The inactive eigenmodes are indicated with a cross. The LO eigenfrequencies at the $\Gamma$ point are indicated by the horizontal lines.

Despite having the same cubic perovskite crystal structure, the $SrTiO_3$ eigenvectors are distinctly different from those of $SrGeO_3$ and $BaSnO_3$. For example, the displacements of the $TO_1$ mode depend on the electronic structure and whether the $B^{4+}$ cation valence orbitals are comprised of Ge 4s (Sn 5s) or Ti 3d orbitals. In $SrGeO_3$ ($BaSnO_3$), the $Ge^{4+}$ ($Sn^{4+}$) cation is displaced along the $Ge^{4+}$—$O_{||}^{2-}$ ($Sn^{4+}$—$O_{||}^{2-}$) bond axis and as part of the $GeO_6$ ($SnO_6$) octahedron in the same direction as the $O_{||}^{2-}$ and $O_{\perp}^{2-}$ anions (Fig. 4b). This is in contrast to the $TO_1$ mode in $SrTiO_3$, for which the $Ti^{4+}$ cations and the $O_{||}^{2-}$ anions of the $TiO_6$ octahedron are displaced in opposite directions (Fig. 4c). The variation of the $Ti^{4+}$—$O_{||}^{2-}$ bond length compresses the bonding O 2p orbital and increases its d character.[56] The dynamic charge transfer from $O_{||}^{2-}$ to $Ti^{4+}$ ions results in anomalously large Born effective charges $e^*_{O_{||}}$ and $e^*_{Ti}$ (Table 4). Consequently, the $TO_1$ mode polarization in $SrTiO_3$ is excessively large, which gives rise to a pronounced TO-LO splitting and a strikingly large $LO_3$ polarization.

Table 4: Calculated Born effective charge $e^*$ (in units of the electron charge $e$) of ions in cubic $SrGeO_3$, and $BaSnO_3$ and $SrTiO_3$ from density functional pertubation theory.

| Material | $e^*_{\mathrm{A}}$ | $e^*_{\mathrm{B}}$ | $e^*_{\mathrm{O}_{\|\|}}$ | $e^*_{\mathrm{O}_{\perp}}$ |
|---|---|---|---|---|
| $SrGeO_3$ | 2.6 | 4.5 | -3.7 | -1.7 |
| $BaSnO_3$ | 2.7 | 4.4 | -3.3 | -1.9 |
| $SrTiO_3$ | 2.5 | 7.7 | -6.1 | -2.1 |

We do not consider the triply degenerate, IR inactive eigenmode any further, which is characterized by the displacement of two $\mathrm{O}_{\perp}^{2-}$ ions in opposite directions without a net ionic dipole moment (Fig. S3). The calculation data of eigenmodes at various points in the phonon band structures is provided as supplementary material for interactive visualization of ion displacements using the website of Miranda.[57] Our results for the $BaSnO_3$ and $SrTiO_3$ eigenmodes agree with a recent phonon calculation,[58] and the analyses of experimental dispersion data obtained from single crystals,[25,59] respectively.

## V. LO PHONON SCATTERING

### A. Determination of the scattering limited electron mobility

In all cubic perovskites of this study, the $LO_1$ phonon polarization is negligibly weak. Thus, the electron-phonon interaction is governed by the large polarization of the $LO_2$ and $LO_3$ phonons. We describe the electron-phonon interaction quantitatively by polaron coupling constants $\alpha_\nu$ according to[53,60]

$$\alpha_\nu = \frac{1}{8\pi\varepsilon_0}\frac{e^2}{r_p\hbar\omega_{\mathrm{l}_\nu}}\left(\frac{1}{\varepsilon_\infty}-\frac{1}{\varepsilon_\mathrm{s}}\right)\frac{\varepsilon_{\mathrm{l}_\nu}}{\sum_\nu \varepsilon_{\mathrm{l}_\nu}} \tag{5}$$

where $e$ is the electron charge and $r_p = \left(\hbar/2m_\mathrm{e}^*\omega_{\mathrm{l}_\nu}\right)^{1/2}$ is the polaron radius. The polaron binding energy $\alpha_\nu\hbar\omega_{\mathrm{l}_\nu}$ as derived by Fröhlich[60] is shown to be valid for weak coupling interactions ($\alpha_\nu < 10$). Fröhlich's model was extended to describe the coupling constants of materials having more than one optical phonon mode, by weighting them according to their phonon oscillator strength, as given by the term $\varepsilon_{\mathrm{l}_\nu}/\sum_\nu \varepsilon_{\mathrm{l}_\nu}$.[53] The electron-phonon coupling is weak when I) the effective mass $m_\mathrm{e}^*$ is small, II) the high-frequency dielectric constant $\varepsilon_\infty$ is large to promote screening of the polarization, and III) the oscillator strength $\varepsilon_{\mathrm{l}_\nu}$ of the LO phonon mode is weak. The polaron coupling constants $\alpha_\nu$ for all LO phonon modes are summarized in Table 2.

The LO phonon energy $\hbar\omega_{\mathrm{l}_\nu}$ and the coupling parameter $\alpha_\nu$ determine the relaxation time $\tau_\nu$ of momentum loss by carrier scattering due to electron-LO phonon interactions. It is calculated for each of the LO phonon modes according to[61]

$$\tau_\nu = \frac{1}{2\alpha_\nu\omega_{\mathrm{l}_\nu}}\left(\frac{m_\mathrm{e}^*}{m_\mathrm{p}}\right)^2 f(\alpha_\nu)\left[\exp\left(\frac{\hbar\omega_{\mathrm{l}_\nu}}{k_\mathrm{B}T}\right) - 1\right] \tag{6}$$

where $k_\mathrm{B}$ is the Boltzmann constant, $T$ is the absolute temperature and $f(\alpha_\nu)$ is a dimensionless function, slowly increasing from 1.0 to 1.2 for $0 < \alpha_\nu < 3$. The polaron mass $m_\mathrm{p} = (1 + \alpha_\nu/6)m_\mathrm{e}^*$[44] describes the larger effective mass of an electron which creates a local polarization due to the Coulombic interaction with the ions of the crystal. Eq. (6) for the relaxation time is valid for temperatures smaller than the Debye temperature of the phonon mode ($T \ll \hbar\omega_{\mathrm{l}_\nu}/k_\mathrm{B}$). This condition is satisfied at RT for all considered phonon modes with the exception of the soft $LO_1$ modes, which in any case can be neglected due to their negligibly small polarization.

The relaxation time $\tau_\nu$ is largely dependent on the LO phonon energy $\hbar\omega_{\mathrm{l}_\nu}$ and the oscillator strength $\varepsilon_{\mathrm{l}_\nu}$ as shown in Fig. 5. In general, a hard LO phonon mode exhibits a large coupling parameter, but the thermal occupation at RT, as given by the term $\exp\left(-\hbar\omega_{\mathrm{l}_\nu}/k_\mathrm{B}T\right)$, is small as compared to soft modes. The smallest relaxation time $\tau_\nu$ among all LO phonon modes in each material is crucial for the effective relaxation time $\tau_\mathrm{LO}$, following Mathiessen's rule[53]

$$\tau_\mathrm{LO} = \left(\sum_\nu \frac{1}{\tau_\nu}\right)^{-1} \tag{7}$$

The electron mobility $\mu_\mathrm{LO}$ is determined from the ratio of the effective scattering relaxation time $\tau_\mathrm{LO}$ and the effective mass $m_\mathrm{e}^*$, according to[44]

$$\mu_\mathrm{LO} = \frac{e\,\tau_\mathrm{LO}}{m_\mathrm{e}^*} \tag{8}$$

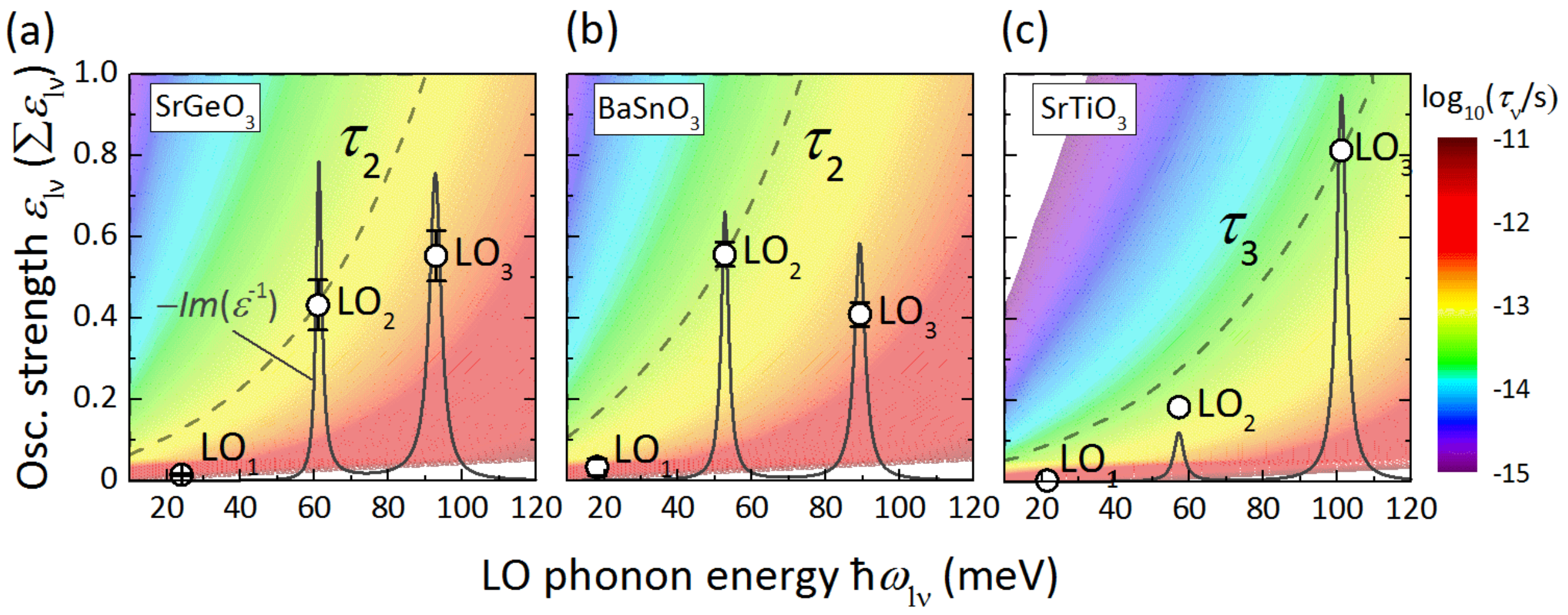

Fig. 5: Relaxation time of carrier scattering $\tau_\nu$ at T = 298 K by LO phonons as a function of the phonon energy $\hbar\omega_{l_\nu}$ and oscillator strength $\varepsilon_{l_\nu}$ for cubic (a) $SrGeO_3$, (b) $BaSnO_3$ and (c) $SrTiO_3$, according to Eqs. (5) and (6). The smallest relaxation time among the three LO phonon modes, which is the most crucial for the intrinsic electron mobility, is indicated by the dashed line. The LO peak heigths in the loss function $-\mathrm{Im}(\varepsilon^{-1})$ indicate the magnitude of polarization, but it must be considered that the peak broading due to damping ($\gamma_\mu > 0$) is taken into account.

Table 5: Lattice parameter $a$, optical and static dielectric constants $\varepsilon_\infty$ and $\varepsilon_s$, and effective mass $m_e^*$ of cubic $SrGeO_3$, $BaSnO_3$ and $SrTiO_3$. The effective RT relaxation time $\tau_{LO}^{RT}$ and RT electron mobility $\mu_{LO}^{RT}$ is calculated according to Eq. (7) and (8), respectively.

| Material | Lattice parameter | Dielectric constants | | Effective mass | Effective RT relaxation time | RT electron mobility |
|---|---|---|---|---|---|---|
| | $a$ | $\varepsilon_\infty$ | $\varepsilon_s$ | $m_e^*$ | $\tau_{LO}^{RT}$ | $\mu_{LO}^{RT}$ |
| | (Å) | - | - | ($m_e$) | ($10^{-14}$ s) | ($cm^2V^{-1}s^{-1}$) |
| $SrGeO_3$ | 3.798[a)] | $4.2 \pm 0.2$ | $18 \pm 1$ | 0.22 | $4.9 \pm 0.4$ | $3.9 \pm 0.5 \times 10^2$ |
| $BaSnO_3$ | 4.117[b)] | 4.3[d)] | $20 \pm 2$[d)] | $0.19 \pm 0.02$[f)] | $3.2 \pm 0.3$ | $3.0 \pm 0.6 \times 10^2$ |
| $SrTiO_3$ | 3.905[c)] | 5.2[e)] | 310[e)] | 1.8[g)] | 1.7 | $2 \times 10$ |

a)Reference[31]; b)Reference[62]; c)Reference[63]; d)References[16,17]; e)Reference[25]; f)Reference[14]; g)Reference[64]

We calculate the intrinsic RT mobilities for cubic $SrGeO_3$, $BaSnO_3$, and $SrTiO_3$ of $3.9\times10^2$, $3.0\times10^2$ and $2\times10$ $cm^2V^{-1}s^{-1}$, respectively (Tables 2 and 5).

The uncertainties in the determination of the high-frequency dielectric constant $\varepsilon_\infty$, the electron effective mass $m_e^*$ and the LO phonon oscillator strength $\varepsilon_{1_\nu}$ are the critical parameters for the error in determining the electron mobility. The combined evaluation of the IR reflectance and Raman scattering of cubic $SrGeO_3$ presented in this study allows for a very reliable determination of both, the phonon energies and oscillator strengths. Therefore, the resulting error of the mobility calculation from these parameters is comparably small. Despite the experimental uncertainty, the present method for calculating the electron mobility is powerful from the viewpoint of providing a very intuitive understanding of how individual phonon modes contribute to the effective relaxation time. In addition, it allows one to circumvent the large computational effort of DFT calculations. Furthermore, theoretical papers on DFT calculations of the phonon scattering-limited mobility often adopt experimental values such as the effective mass, dielectric constants and phonon energies as input parameters, without consideration of the experimental error. Thus, the rather generous error limits of the present calculation results shall not be surprising.

### B. Differences between the phonon spectra and implications for the scattering relaxation time

#### 1. *$SrGeO_3$ and $BaSnO_3$*

In both materials, $SrGeO_3$ and $BaSnO_3$, the $LO_2$ phonon mode is the most crucial one for the scattering (Fig. 5). This mode is described by a vibration of the $Ge^{4+}$ ($Sn^{4+}$) cation against the two $O_\perp^{2-}$ anions in the equatorial positions of the $GeO_6$ ($SnO_6$) octahedron, while the displacement vectors of $Sr^{2+}$ ($Ba^{2+}$) and $O_{||}^{2-}$ are comparably small (cf. Fig. 4b). Considering only the displacement of $Ge^{4+}$ ($Sn^{4+}$) and $O_\perp^{2-}$ ion sublattices, and regarding other ions are fixed on their equilibrium lattice sites, the phonon eigenfrequency is proportional to the square root of the reduced mass of the $Ge^{4+}$ ($Sn^{4+}$) and $2\times O_\perp^{2-}$ ions[44]

$$\omega_{t_\mu}^2 = 2k_\mu\left(\frac{1}{M_\alpha}+\frac{1}{M_\beta}\right) \tag{9}$$

as follows from the analogy to the case of a diatomic crystal. Here, $M_\alpha$ and $M_\beta$ denote the mass of the ions in the two ion sublattices and $k_\mu$ is the force constant. Thus, it is concluded that the $TO_2$ and $LO_2$ eigenfrequencies in $SrGeO_3$ (291 and 493 $cm^{-1}$) are increased with respect to the corresponding modes in $BaSnO_3$ (246 and 425 $cm^{-1}$) due to the smaller atomic mass of Ge (72.6 u) as compared to Sn (118.7 u). The $TO_3$ eigenfrequencies in $SrGeO_3$ (634 $cm^{-1}$) and $BaSnO_3$ (629 $cm^{-1}$) are nearly identical, since they are mainly described by a vibration of only $\mathrm{O}_{||}^{2-}$ and $\mathrm{O}_{\perp}^{2-}$ anions. As is seen from the expansion coefficient matrix to obtain the LO phonon oscillator strengths (Eq. (4) and Table 3), the increased $TO_2$/$LO_2$ eigenfrequencies in $SrGeO_3$ result in a stronger coupling between the $LO_2$–$TO_3$ modes and a weaker coupling between the $TO_3$–$LO_3$ modes as compared to the corresponding ones in $BaSnO_3$. Thus, the relative $LO_2$ oscillator strength $\varepsilon_{\mathrm{l}_2}/\sum_\nu \varepsilon_{\mathrm{l}_\nu}$ in $SrGeO_3$ (1.2/3.2) is smaller as compared to that of BaSnO3 (1.9/3.4), resulting in a somewhat longer RT scattering relaxation time $\tau_2^{\mathrm{RT}}$ (Tables 2 and 5).

### *2. $SrTiO_3$*

Among the $SrTiO_3$ LO phonons, the hard $LO_3$ mode exhibits the largest polarization and it is the most crucial one for the intrinsic mobility (Fig. 5c). However, the hard $LO_3$ mode is the least thermally occupied mode at RT among all three LO modes (cf. Eq. (6)). Considering the large $SrTiO_3$ effective mass of 1.8 $m_e$, the resulting 17-ps scattering relaxation time is nevertheless remarkably long, and only by a factor of 2-3 shorter than that obtained for $SrGeO_3$ and $BaSnO_3$ (cf. Table 5).

## C. Implications for the design of high-RT electron mobility semiconductors

The results of our analysis provide an important guideline for the design of high-RT mobility semiconductors. The most promising candidate materials shall combine both properties, a small effective mass *and* a comparably large polarization for the hardest LO mode. A large static dielectric constant is indicative of a strongly polar TO phonon mode which in turn yields an increased polarization of hard LO modes. Transition-metal oxides commonly exhibit a large static dielectric constant due to the dynamic charge transfer between ions and increased Born effective charges, but the compromise is a large effective mass, since d-orbitals are generally more localized than s-orbitals. Thus, it would be very exciting to explore transition-metal based compound semiconductors which, in analogy to cubic $SrGeO_3$ and $BaSnO_3$, show a nonbonding character of the cation-derived CB, and consequently, a reduced effective mass. The recently synthesized tetragonal ZrOS semiconductor with a dispersive CB derived from a Zr 3d non-bonding state and small 0.36 $m_e$ effective mass proves that such materials design is feasible,[65] providing very interesting candidates for future investigation.

### D. Comparison of the calculated electron mobility with experimental results

The calculated RT mobility on the order of $3.0 \pm 0.6\times10^2$ $cm^2V^{-1}s^{-1}$ for $BaSnO_3$ is in very good agreement with experimental values determined by Hall effect measurements of La-doped single crystals (220–320 $cm^2V^{-1}s^{-1}$).[6,28] Furthermore, the calculation result of $2\times10$ $cm^2V^{-1}s^{-1}$ for $SrTiO_3$ agrees well with the measured Hall mobility of about 5 $cm^2V^{-1}s^{-1}$ determined for La-doped $SrTiO_3$ epitaxial thin films.[22]

To determine the electrical transport properties in cubic $SrGeO_3$ experimentally, we fit the IR reflectance spectrum of a mirror-polished 2 at.% La-doped $SrGeO_3$ polycrystalline pellet (Fig. 6). At such high doping level, the phonon spectrum is screened due to the large electronic contribution of free carriers to the dielectric function. Thus, the fitting is performed by only using the Drude model,[44]

$$\varepsilon_D(\omega) = \varepsilon_\infty - \frac{\varepsilon_\infty \omega_{\mathrm{p}}^2}{\omega^2 + i\omega/\tau_{\mathrm{p}}} \tag{10}$$

where $\omega_p$ denotes the unscreened plasma frequency and $\tau_{\mathrm{p}}$ denotes the carrier scattering relaxation time. We determine the optical mobility of $42 \pm 3$ $cm^2V^{-1}s^{-1}$ (Eq. (8)), the carrier concentration of $1.3 \pm 0.1\times10^{20}$ $cm^{-3}$ (Eq. S(1)) and the optical resistivity of $1.1 \pm 0.1\times10^{-3}$ $\Omega cm^{-1}$ (Eq. S(2)). The reduced optical mobility as compared to the calculation result may indicate the presence of defects, including the possibility of Ge vacancy formation or a reduction of $Ge^{4+}$ to $Ge^{2+}$. The Hall-effect measurement of substitutionally doped $SrGeO_3$ single crystals will be the subject of further investigation, when crystals of larger size can be grown and characterized.

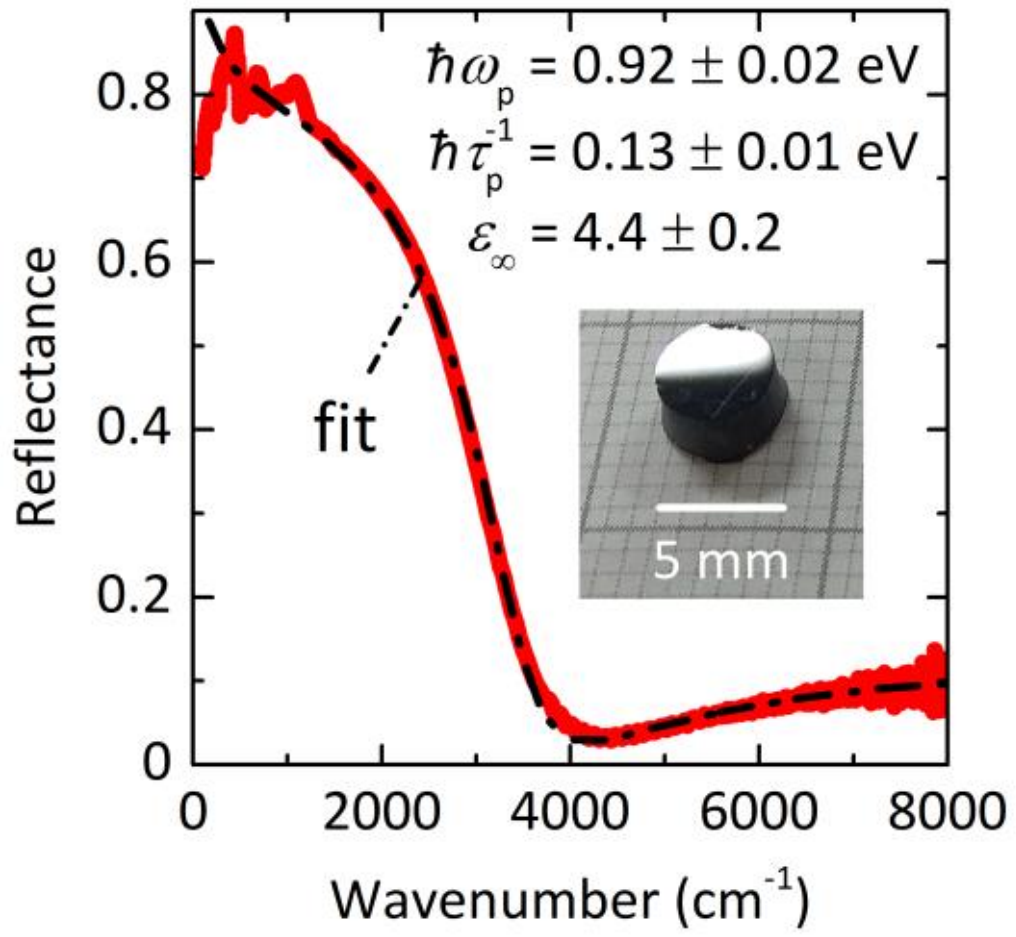

Fig. 6: The IR reflectance spectrum of a mirror-polished, 2 at.% La-doped cubic $SrGeO_3$ ($La_{0.02}Sr_{0.98}GeO_3$) polycrystalline pellet.

## VI. CONCLUSION

In summary, we have successfully synthesized cubic $SrGeO_3$ perovskite single crystals using the high-pressure flux growth. The optical phonon spectrum is determined from IR optical reflectance and Raman analyses to evaluate the electron transport governed by LO phonon scattering. Employing the classical phonon theory and a combined experimental-theoretical approach, we analysed the individual eigenmodes to gain a fundamental understanding of the similarities and differences in the phonon spectra for the cubic perovskites $SrGeO_3$, $BaSnO_3$ and $SrTiO_3$. Cubic $SrGeO_3$ exhibits a calculated RT mobility in the order of $3.9\times10^2$ $cm^2V^{-1}s^{-1}$, demonstrating great promise as high-performance transparent conducting oxide. We suggest a universal guideline for the exploration of new high RT mobility semiconductors. The combined properties of a small effective mass *and* a large polarization of the hardest LO mode is in favor of supporting an exceptionally high RT mobility despite the electron-phonon interactions.

## Acknowledgements

We are thankful to Dr T. Stanislavchuk from New Jersey Institute of Technology for sharing the $BaSnO_3$ IR ellipsometry data. The work at Tokyo Institute of Technology was supported by the MEXT Element Strategy Initiative to Form Core Research Center (JPMXP0112101001). C.A.N. acknowledges the support through a fellowship granted by the German Research Foundation (DFG) for proposal NI1834. T. Katase was supported by PRESTO, Japan Science and Technology Agency (JPMJPR16R1) and Grant-in-Aid for Scientific Scientists B (19H02425) from JSPS.

# Supporting Information for

# Phonon scattering limited mobility in the representative cubic perovskite semiconductors $SrGeO_3$, $BaSnO_3$ and $SrTiO_3$

Christian A. Niedermeier,[1,*] Yu Kumagai,[2] Keisuke Ide,[1] Takayoshi Katase,[1] Fumiyasu Oba,[1,2,] Hideo Hosono,[1,2] and Toshio Kamiya[1,2]

**Corresponding author: c-niedermeier@mces.titech.ac.jp*

[1]Laboratory for Materials and Structures, Tokyo Institute of Technology, Yokohama 226-8503, Japan

[2]Materials Research Center for Element Strategy, Tokyo Institute of Technology, Yokohama 226-8503, Japan

## I. Cubic $SrGeO_3$ single crystal quality

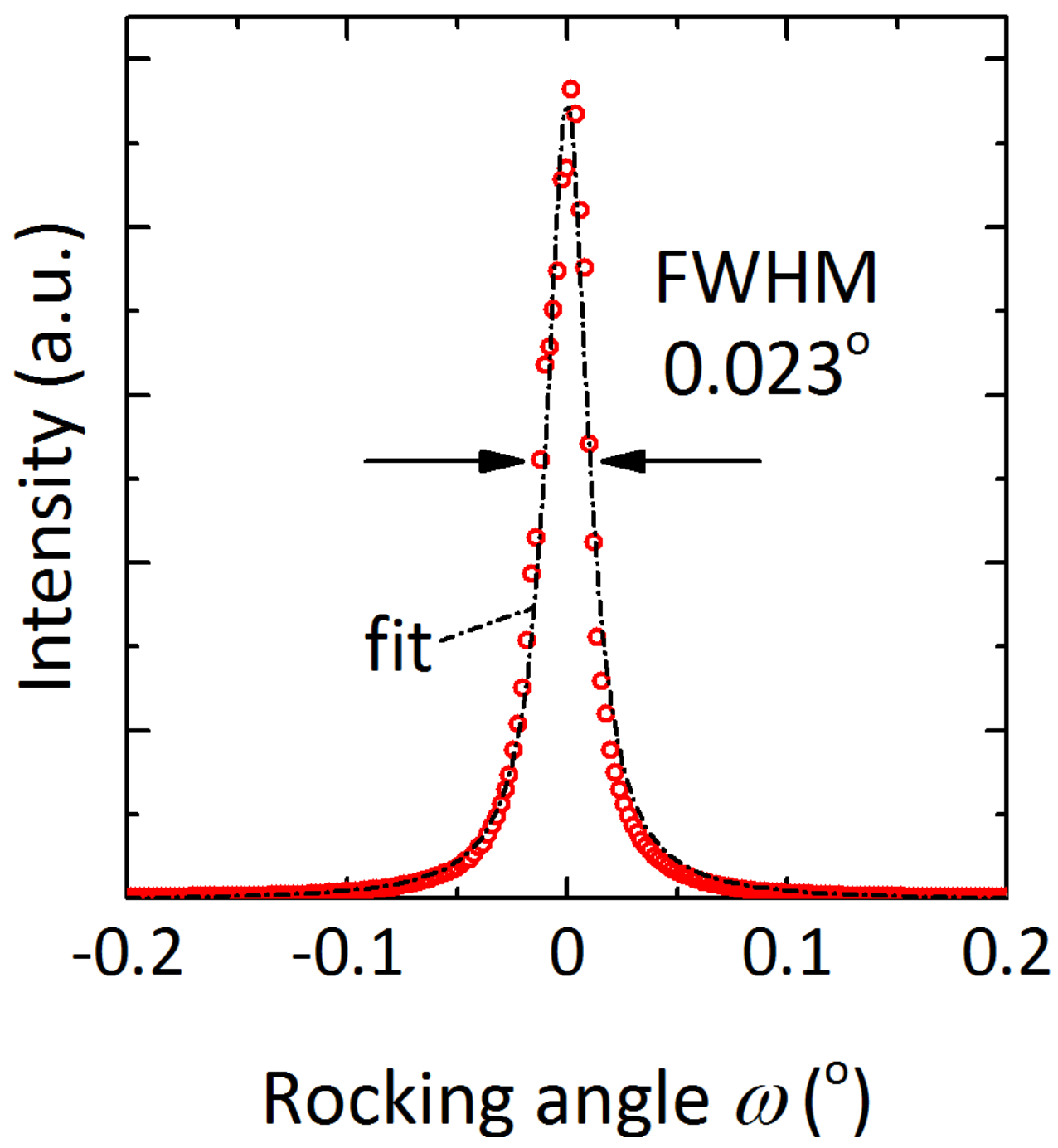

Fig. S1: The rocking curve of the 200 XRD peak of a cubic $SrGeO_3$ single crystal.

## II. Fitting of the $BaSnO_3$ IR ellipsometry spectrum

The frequency and oscillator strength of the TO phonon modes in $BaSnO_3$ at 135 ($TO_1$), 246 ($TO_2$) and 630 cm$^{-1}$ ($TO_3$) were determined from the IR reflection ellipsometry spectrum of a single crystal, obtained by Stanislavchuk (Fig. S2).[1] We re-evaluate Stanislavchuk's dispersion model, since the reported $BaSnO_3$ high-frequency dielectric constant ($\varepsilon_\infty$= 3.3) would underestimate the real part of the dielectric constant at low energies, $\varepsilon_1$(0.75 eV) = 4.3. The first-order Raman peak at 724 cm$^{-1}$ assigned to the $LO_3$ phonon[1] is used as an additional constraint while allowing for some fitting variance. The real part $\varepsilon_1$ and the imaginary part $\varepsilon_2$ of $BaSnO_3$ dielectric function were fitted using Eq. (2) and employing the RefFit program.[2]

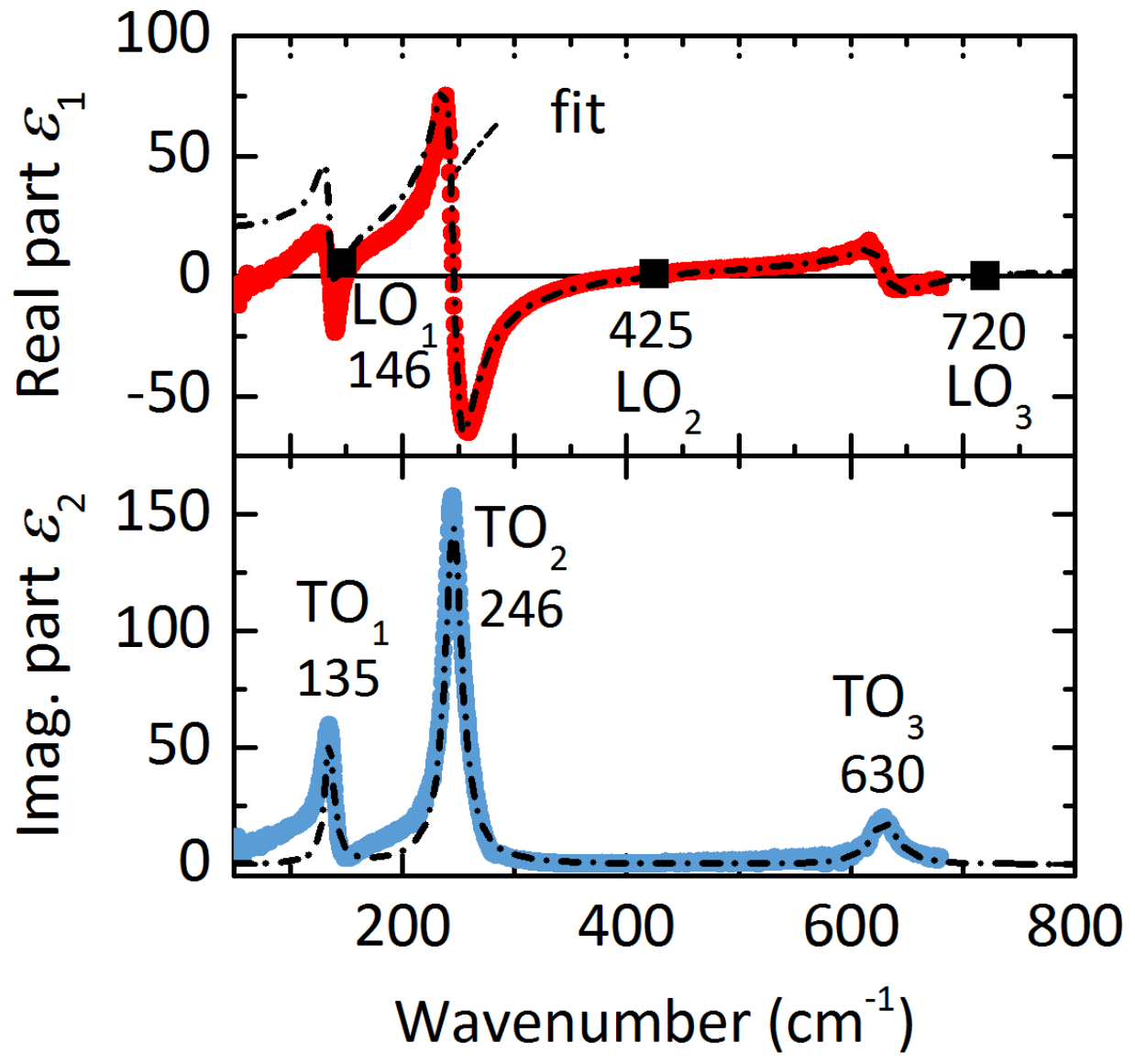

Fig. S2: The real part $\varepsilon_1$ and the imaginary part $\varepsilon_2$ of the $BaSnO_3$ dielectric function recorded using IR reflection spectroscopic ellipsometry of a single crystal, reported by Stanislavchuk.[1] Dashed lines indicate our fitting result using the three Lorentz oscillator model of Eq. (2) and the dispersion parameters given in Table 1.

It has been noted that due to the small 3×2×1 mm$^3$ $BaSnO_3$ crystal size, measurement artefacts result in an artificial decrease (increase) of the real part $\varepsilon_1$ (imaginary part $\varepsilon_2$) of the dielectric function below a frequency of about 150 cm$^{-1}$. While the $TO_1$ mode eigenfrequency $\omega_{t_1}$ = 135 $cm^{-1}$ of is determined accurately, the corresponding oscillator strength $\varepsilon_{t_1}$ = 3.8 ± 2 shows some uncertainty.

## III. The IR inactive phonon mode in cubic perovskites

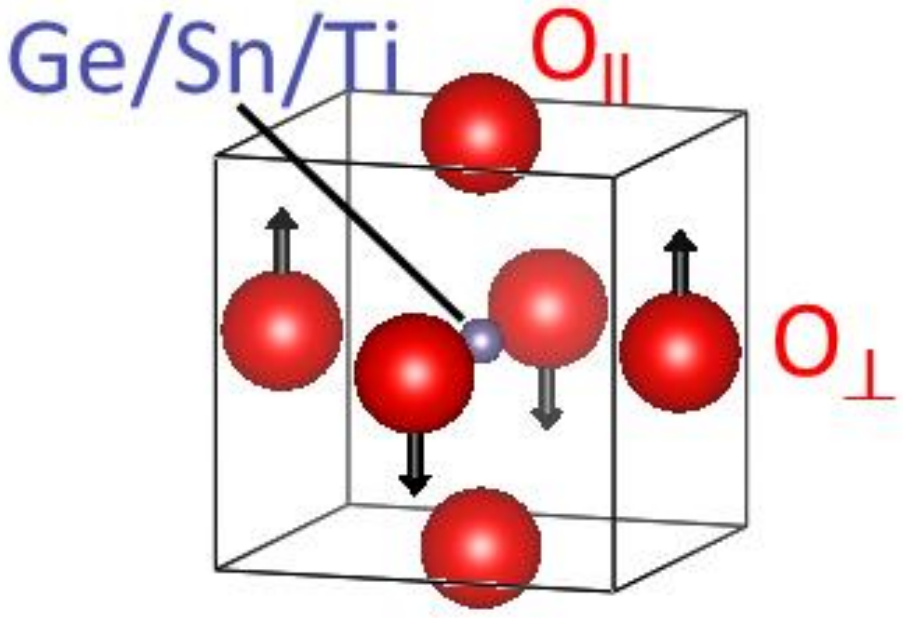

Fig. S3: The IR inactive phonon eigenmode in cubic $SrGeO_3$ (267 $cm^{-1}$), $BaSnO_3$ (214 $cm^{-1}$) and $SrTiO_3$ (223 $cm^{-1}$). The eigenfrequencies are determined by the phonon calculation.

## IV. Determination of electrical properties from the Drude model parameters

The carrier concentration is given by[3]

$$n_\mathrm{e} = \frac{m_\mathrm{e}^* \varepsilon_0 \varepsilon_\infty \omega_\mathrm{p}^2}{e^2} \quad \text{(S1)}$$

and the optical resistivity is given by[3]

$$\rho = \frac{m_\mathrm{e}^*}{\varepsilon_0 \varepsilon_\infty \tau_\mathrm{p} \omega_\mathrm{p}^2} \quad \text{(S2)}$$